\documentclass[useAMS,usenatbib]{mn2e}
\usepackage{psfig}
\usepackage{amsmath}
\usepackage{amssymb}
\def \AAO {AA$\Omega$}
\def \LL {L$\times$L}
\def \LR {L$\times$R}

\def \LM {L$\times$M}

\def \rs {$^{\rm r*}$}
\def \ts {$^{26*}$}
\def \AAOts {AA$\Omega^{26*}$}

\title[Radio clustering evolution]{Evolution in the clustering
  strength of radio galaxies}

\author[Fine et al.]
       {S. Fine$^1$\thanks{stephen.fine@durham.ac.uk},
         T. Shanks$^1$, N. Nikoloudakis$^1$, U. Sawangwit$^1$   \\
$^1$Department of Physics, Durham University, South Road, Durham DH1
         3LE, UK \\
}

\begin{document}

\maketitle

\begin{abstract}

We cross match the NVSS and FIRST surveys with three large photometric
catalogues of 
luminous red galaxies (LRGs) to define radio-loud samples. These
have median redshifts 0.35, 0.55 and 0.68 and,
by matching rest-frame optical and radio properties, we construct
uniform samples across the three surveys. This paper is concerned with
the clustering properties of these samples derived from the angular
correlation function. The primary aim is to
characterise any evolution in the clustering amplitude of radio
galaxies bellow $z\sim0.68$.

We find no evidence for evolution in the large-scale
($\sim1-50\,h^{-1}$\,Mpc) clustering amplitude.
Our radio galaxy
autocorrelations are consistent with previous findings indicating
little-to-no evolution in the redshift range 0.68 to 0 ($\sim6$\,Gyr of
time). We also cross correlate radio galaxies
with the parent LRG samples to increase the precision of our results
and again find no evidence for evolution in comoving coordinates.
% Assuming $r_0\propto(1+z)^\alpha$ we find
%$\alpha=0.20\pm0.30$. Furthermore, we cross correlate the radio galaxies
%with the pairent LRG samples to increase the precision of our results
%and again find no evidence for evolution in comoving coordinates.
%($\alpha=0.23\pm0.21$).
Our results are inconsistent with a long-lived
model for the clustering evolution that assumes radio sources randomly
sample the LRG population. A model where the halo
mass is constant with redshift is consistent with the
data. This is similar to QSOs that have
clustering amplitudes consistent with a single halo mass at all
redshifts.
Given that the brightest radio sources show stronger evolution in space
density compared to fainter radio sources we restrict our samples to
include only objects with $L>10^{26}$\,W/Hz and repeat the analysis. Again
we find no evidence for evolution in the comoving correlation amplitude.
These radio sources appear to inhabit the same mass halos as fainter
radio galaxies ($\sim9\times10^{13}h^{-1}$\,M$_\odot$). These halos
are $\sim$twice as massive as those of the general LRG 
population ($\sim4\times10^{13}h^{-1}$\,M$_\odot$) and 
$\sim30$ times as massive as optical
AGN/QSOs ($\sim3\times10^{12}h^{-1}$\,M$_\odot$). 

\end{abstract}


\section{Introduction}

Radio galaxies have long been used to study large scale
structure as they are highly biased indicators of the mass
distribution and are naturally found in the high redshift ($z\sim1$)
Universe.
%Radio sources are hosted by galaxies and relating the high
%bias of radio sources to their host populations gives insight into the
%triggering mechanisums
Understanding the observed bias of radio galaxies requires relating
their clustering to that of their host populations.
Furthermore, any differential evolution in their clustering
with respect to their hosts gives insight into the nature
of radio galaxies and the mechanisms that trigger them.

%of radio source population we
%must understand how the bias of radio galaxies is related to their
%host galaxy populations and how this relationship evolves.

% However, in order to be useful probes it is vital to
%understand how radio galaxies trace the underlying dark matter
%distribution and how this evolves with redshift \citep{bra05}.

The typical host galaxy of a radio source depends strongly on its
luminosity. At lower radio luminosities ($L\lesssim10^{23}$\,W/Hz) the
population is almost exclusively made up of starforming galaxies with
related diffuse radio emission, above $L\sim10^{23}$\,W/Hz almost
all radio sources 
are associated with active galactic nuclei (AGN). The hosts
of the brightest radio galaxies have strong emission lines
(e.g. \citealt{h+l79,mcc93}) while lower-luminosity radio AGN are hosted
by massive ellipticals with little-to-no optical line emission. This
transition implies a cutoff luminosity between objects with and
without emission lines that \citet{joh08} found to be around
$\sim10^{26}$\,W/Hz. This luminosity is close to the
traditional cutoff between morphologically classified FRI and FRII
objects \citep{f+r74}. Furthermore, while the bright end of the radio
luminosity 
function evolves strongly towards higher space densities at high $z$
\citep{lon66,d+p90}, fainter radio AGN ($L\lesssim10^{26}$\,W/Hz) show
much less evidence for evolution \citep{c+j04,sad07}.
This evolution in space density is roughly matched by the typical
hosts of radio AGN.
Massive elliptical galaxies (luminous red galaxies; LRGs), that host
lower luminosity radio AGN, show little evolution in number
density for $z<1$ \citep{bel04,wak06,bro07}. Higher luminosity radio sources
are more associated with optical/Xray AGN \citep{mcc93} that show
strong evolution out to $z\sim2.5$ \citep{sch68,croom09b}.

Radio sources are
strongly biased tracers of the underlying mass distribution of the
Universe \citep{ymp89,p+n91,b+w02,mag04,bra05}. Since the advent of
deep all sky surveys (e.g. FIRST, NVSS, SUMSS, WENSS)
%\citep{bec00,con98,mau03,ren97}
there have been major leaps in the
ability to describe the environments of radio galaxies. These
milli-Jansky surveys
sample the luminosity range $\sim10^{23}$ to $10^{26}$\,W/Hz at
moderate redshifts, where radio sources are found almost exclusively
in LRGs. LRGs are known to be a strongly biased tracer of the matter
density, and their bias increases with the mass/luminosity of the
LRG.

Radio sources tend to be found in the most massive elliptical
galaxies and more massive galaxies are more likely to host them
\citep{bes05,m+s07}. Recent studies have
compared the clustering of radio galaxies to that of 
radio-quiet galaxies that have been matched in their optical properties
to the radio sample. Most find that radio galaxies are
significantly more clustered than optically identical samples of
quiescent galaxies
\citep{wak08a,man09,don09}, although see \citet{hic09} for a counter
example we will discuss later.
%\citet{hic09} found no evidence for this,
%and suggested that differences in the ratio of radio powed to stellar
%mass of the galaxies could explain their findings.
The indication is that
the environment of a galaxy contributes to the probability of it hosting
a radio source.

In this paper we are primarily concerned with how the clustering of
radio sources evolves for $z\lesssim0.7$. In this redshift interval
the clustering amplitude of LRGs is approximately constant
\citep{wak08b,saw09}. However, the clustering amplitude of
quasars evolves such that, when models of gravitational
collapse are assumed, the implied dark halo mass for quasars is
approximately constant \citep{croom05,ros09}. We will evaluate
the clustering amplitude of radio LRGs in a series of samples from
$z\sim0.68$ to 0.35 to investigate any evolution in their
clustering. This will be done both for a sample of medium luminosity
($L\sim10^{24.7}$\,W/Hz) radio galaxies, and a high luminosity subsample
($L>10^{26}$\,W/Hz).

In section~\ref{sec:data} we introduce the data that will be
used in this work, section ~\ref{sec:rad_match} describes the
algorithm we develop for identifying radio LRGs,
section~\ref{sec:cor_anal} describes the correlation analysis to be
used, section~\ref{sec:rlc} compares the clustering of radio-loud and
quiet LRGs, section~\ref{sec:evol} then looks at the evolution of
radio LRGs and in section~\ref{sec:Levol} we investigate the
evolution of the very brightest radio LRGs. Throughout this paper we
assume a flat $(\Omega_{\rm m},\Omega_{\Lambda})=(0.3,0.7)$,
$H_{0}=70\,{\rm km\,s}^{-1}\,{\rm Mpc}^{-1}$ cosmology. All radio flux
densities and luminosities are at 1.4\,GHz unless otherwise stated and
when estimating radio luminosities throughout this paper $k$-corrections
have been performed assuming a continuum shape of
$S_\nu\propto\nu^{-0.7}$.

\section{Data} \label{sec:data}

%This paper is concerned with the radio properties of three photometric
%samples of LRGs.

The LRG samples used in this paper are originally defined from three
spectroscopic surveys: SDSS, 2SLAQ and \AAO\ \citep{eis01,can06,ros08}.
The LRG selection was refined by \citet{saw09} to create photometric
samples from the SDSS DR5 \citep{yor00,ald07} and to cut down on stellar
contamination. We use
LRG samples defined identically but drawn from the more recent DR7
\citep{aba09}.

% We 
%will refer to them in order of increasing redshift as 
%the SDSS, 2SLAQ and \AAO\ samples, as their selection was defined by
%these spectroscopic samples. Their median redshifts are 0.35,
%0.55 and 0.68.

To identify radio-loud LRGs in our sample we compare our LRG
catalogues with radio source catalogues from the NVSS \citep{con98} and
FIRST \citep{bec00} surveys. Both surveys are carried out at 1.4\,GHz
and the high angular resolution ($\sim$5\,arcsec) and faint flux
limit ($1\sigma\sim$0.15\,mJy) of the FIRST
survey combined with the large-scale sensitivity of the
NVSS make them complementary tools for identifying radio LRGs.
The FIRST survey covers much of the same sky area as the SDSS north
Galactic cap. In the overlap region
there are 110104, 652401 and 799519 objects in the SDSS, 2SLAQ and
\AAO\ LRG samples respectively. See \citet{saw09} for the
redshift distributions, expected contamination and autocorrelation
clustering properties of the LRG samples.

\section{Radio cross matching} \label{sec:rad_match}

%In order to study the radio properties of the LRG samples we
%cross-match the NVSS and FIRST radio source catalogues with our sample to
%obtain a sample of radio-loud LRGs.

Complex radio source morphology
due to extended structures compounded by size-dependent response
effects due to the use of interferometers makes radio source matching
more involved
than simply matching sky positions. The most accurate and precise
method for cross-matching has always been manual inspection
(e.g. \citealt{sad07}). However, the size of our LRG samples forces
us to define an automated cross-matching procedure that does
not require visual inspection of all of the potential radio matches in
the sample.
Automated radio-matching routines have been developed before
(e.g. \citealt{bes05,k+i08}), and even samples defined by manual inspection
use some automated procedures to define a sample of potential matches
that are then visually inspected.

\subsection{Our cross-matching procedure}

\citet{sad07} define two samples of radio LRGs from the 2SLAQ spectroscopic
sample. One is based on manual inspection and the other based on
automatic cross matching to the FIRST survey.
%In addition to producing a catalogue of radio LRGs from the 2SLAQ
%spectroscopic sample, \citet{sad07} also define a radio sample based
%on FIRST matches to the photometric input catalogue to the 2SLAQ LRG
%survey.
We modified their selection criteria to define a radio matching
procedure and use their manually inspected sample to test our
results. We match the LRG and FIRST catalogues within a 30\,arcsec
radius, and the LRG and NVSS catalogues within a 180\,arcsec
radius, retaining all radio matches within the given radius around an
LRG. In this initial cross matching we do not apply any radio flux
limits, these are applied below. Following \citet{sad07} we use the
FIRST matches as our primary 
tool for identifying radio galaxies. We accept matches that meet one
of three criteria:

\begin{itemize}
\item[{\it i})]{
A FIRST match is within 3\,arcsec of the LRG position.}
\item[{\it ii})]{
A FIRST match is within 10\,arcsec of the LRG position and the major
axis of the source is orientated within 25$^\circ$ of the angle to
the LRG position.}
\item[{\it iii})]{
Given more than one radio match, the flux-weighted mean of the FIRST
positions (or any subset thereof) is within 6\,arcsec of the
LRG position.}
\end{itemize}

For any sources that had NVSS matches within 180\,arcsec but not FIRST
matches we applied a similar set of criteria based on \citet{bes05}
to the NVSS sources to define potential radio galaxies. These are:

\begin{itemize}
\item[{\it i})]{
An NVSS match is within 10\,arcsec of the LRG position.}
\item[{\it ii})]{
A galaxy matched to two radio components that is $<10$\,arcsec from the
flux weighted radio centroid, and the radio-galaxy-radio angle is
$>150^\circ$}
\item[{\it iii})]{
Given more than two radio matches, the flux-weighted mean of the NVSS
positions is within 10\,arcsec of the LRG position.}
\end{itemize}

We apply this procedure to the 2SLAQ spectroscopic catalogue and find
428 objects satisfy our selection criteria. Of these, 57 objects do
not appear in the Sadler catalogue, while 20 objects from the Sadler
catalogue are not found. This indicates that we are $\sim5$\,\%\
incomplete with respect to the Sadler catalogue, the majority of
sources missed by our selection criteria have complex structures in
their FIRST images. Of the 57 extra matches 29 can be explained as objects that
had NVSS matches but no corresponding FIRST sources. These objects
were rejected by Sadler et al. to improve the reliability of their
catalogue. We take an approach more similar to \citet{bes05}
including these sources given the criteria above. This indicates that
our routine, applied to the 2SLAQ spectroscopic LRG catalogue, is
$\sim93$\,\% reliable.

We apply this radio cross-matching routine to the SDSS, 2SLAQ and
\AAO\ photometric LRG samples and obtain
%. We find that the cross matching routine identifies
9689, 15236 and 8364 radio LRGs from the three samples. This
corresponds to 8.8, 2.3 and 1.0\,\% of the parent samples. The
considerably higher detection rate for the SDSS sample is due to the
lower redshift and higher absolute (optical) magnitude of the SDSS LRGs 

We perform a simple check of the reliability of our final radio
samples by offsetting the original LRG samples, then rerunning the
cross-matching code on the offset catalogues. We offset the samples by
$\pm0.1^\circ$ in declination (twice the 180\,arcsec maximum
cross-matching radius 
used). The cross-matching code found 171/138, 516/453, 457/499
matches in the offset SDSS, 2SLAQ and \AAO\ samples respectively for the
positive and negative offsets. This corresponds to $\sim$1.5-6\,\%
contamination in our final catalogues similar to the $\sim7$\,\%
value found when comparing to \citet{sad07}.

%\subsection{Flux limits and completeness}

Many of the radio identifications are based solely on matching to
FIRST source catalogues thanks to the angular resolution of the FIRST
survey. However, the FIRST survey's smaller beam resolves out
large-scale structure and does not give accurate radio powers for
these sources. 
We cross match our final radio LRG catalogue with the NVSS within a
180\,arcsec radius. We then identify which NVSS matches are associated
with the galaxy using the criteria above. We sum
these NVSS components to give the radio flux densities used throughout
this paper.
%Fig.~\ref{fig:NS} shows the
%number counts of radio sources in our \AAO\ sample.
The NVSS is
essentially complete above $\sim2.8$\,mJy \citep{con98} and 
%this can be seen as the number
%counts drop off below this level in Fig.~\ref{fig:NS}. 
in our final sample we apply a radio flux density cut of 3\,mJy.

%\begin{figure}
%\centerline{\psfig{file=plot_NS.ps,width=7cm,angle=-90}}
%\caption{LogN - LogS plot for the radio LRGs in the \AAO\ sample. NVSS is incomplete below 2.8\,mJy as can be seen by the drop in counts below that level. The dotted line indicates the 3\,mJy flux limit we impose on our sample.}
%\label{fig:NS}
%\end{figure}

\subsection{Redshift distributions}

\begin{figure}
%\centerline{\psfig{file=plot_nz_SDSS.ps,width=6cm,angle=-90}\psfig{file=plot_nz_2SLAQ.ps,width=6cm,angle=-90}\psfig{file=plot_nz_AAO.ps,width=6cm,angle=-90}}
\centerline{\psfig{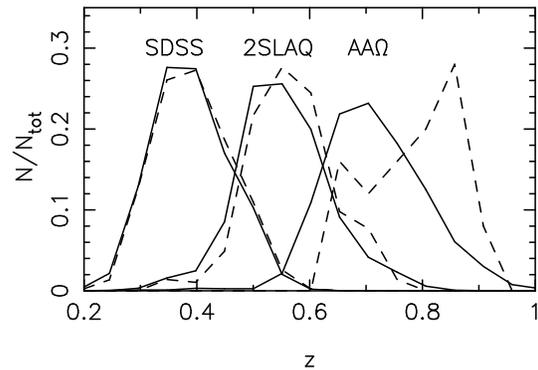}}
\caption{The solid lines shows the redshift distributions of the SDSS,
  2SLAQ and \AAO\ spectroscopic samples (left to right). The dashed
  lines show the 
  redshift distribution of those objects identified as radio
  sources. These are, in general, identical to the parent
  distribution. In the highest redshift \AAO\ sample there is a
  difference, but there are only 17 objects defining this distribution.}
\label{fig:nz}
\end{figure}

The radio-loud subsets of the three LRG samples may not have the same
redshift distribution as their parent sample. In Fig.~\ref{fig:nz} we
plot the redshift distribution of the spectroscopic LRG samples, and those
objects that we identify as radio galaxies from our matching routine.

For the SDSS and 2SLAQ samples it is clear that the radio-detected
subsamples have an identical redshift distribution to their parent
samples. The \AAO\ radio sample shows indications of being at a higher
redshift. However, there are only 17 spectroscopic \AAO\ radio LRGs
and hence the distribution is highly uncertain. For the rest of this
paper we will assume that the radio LRGs have 
the same redshift distribution as their parent sample. None the less, if the
indication of Fig.~\ref{fig:nz} is accurate (that the radio LRGs have
a higher redshift than the parent sample) this will effect our
results. In particular a cross correlation between the radio loud
sample and the parent sample will be reduced if the redshift
distributions do not overlap so well.

\section{Correlation analysis} \label{sec:cor_anal}

To study the clustering of the samples defined above we
calculate the 2-point angular correlation function to measure the
LRG$\times$LRG autocorrelations and the LRG$\times$radio cross correlation.
The 2-point angular correlation function is defined as the relative increase
in pair counts at a given scale over that of a random sample. That is
\begin{equation}
w(\theta) = \frac{N(R)}{N(D)}\frac{DD(\theta)}{DR(\theta)} - 1
\label{equ:2pc}
\end{equation}
where $DD$ gives the number of sample-sample pairs
and $DR$ gives the sample-random pairs at separation $\theta$. The
expression is normalised by the number density of randoms $N(R)$
divided by the density of the sample $N(D)$. In an autocorrelation the
data-data pairs come from the same sample, in a cross correlation they
come from differing samples. Other estimators for $w(\theta)$
\citep{ham93,l+s93} give equivalent results for this analysis.

%Equation~\ref{equ:2pc} gives the basic form of the correlation
%function, but in practise calculations based on equation~\ref{equ:2pc}
%can be affected by edge effects.......

We constructed our random catalogue from the DR7 masks published on
the SDSS website. The full DR7 area was then cut down to the area
covered by FIRST and populated
with random sources with a density twenty times that of the radio LRGs.

\subsection{Error estimation}

Standard Poisson errors are notoriously inaccurate for correlation
functions due to the high level of covariance in the data
(e.g. \citealt{saw09}). More often bootstrap or field-to-field errors
are employed as an independent measure of the precision of the results.
We calculate field-to-field errors by subdividing our sample into nine
declination strips that contain even numbers of objects in our random
catalogues. We calculate the correlation function within each
field and the error is estimated by the rms of the correlation functions
in each field divided by three.

\section{Correlation functions} \label{sec:rlc}

\begin{figure*}
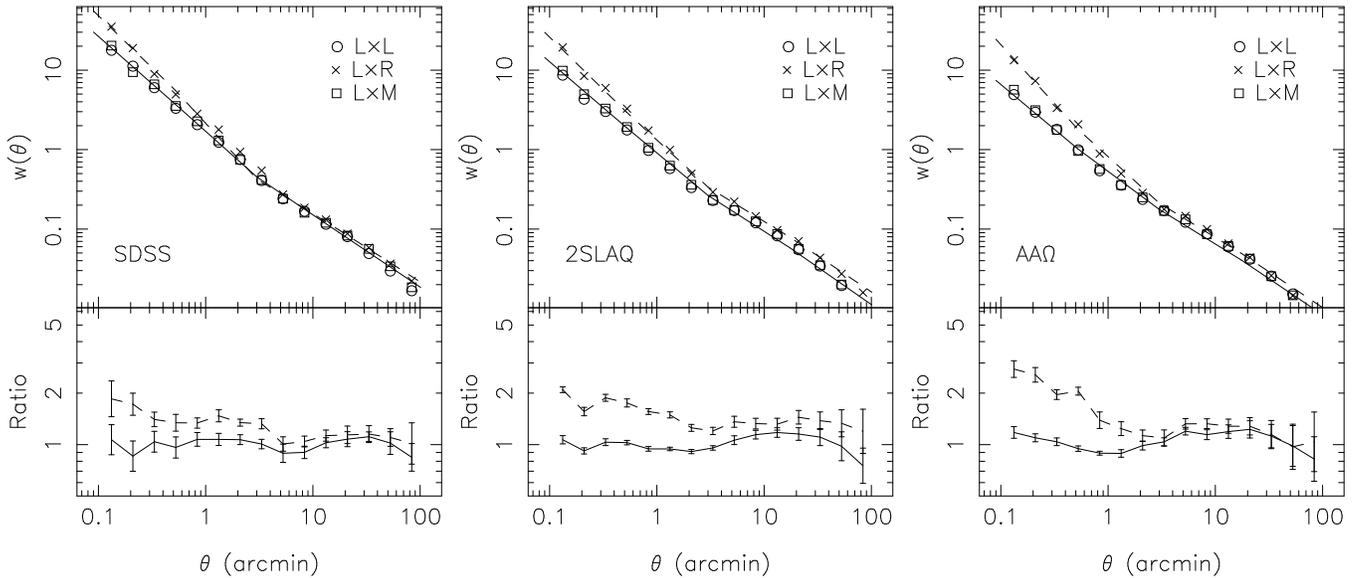

\centerline{\psfig{file=plot_radMatch_SDSS.ps,width=5.8cm}\hspace{0.2cm}\psfig{file=plot_radMatch_2SLAQ.ps,width=5.8cm}\hspace{0.2cm}\psfig{file=plot_radMatch_AAO.ps,width=5.8cm}}
\caption{The \LL\ and \LR\ correlation functions for the SDSS,
  2SLAQ and \AAO\ samples (left to right). Also shown is the \LM\
  cross correlation of the sample matched to the radio samples in
  terms of optical luminosity and colour. The top panels show
  the correlation functions along with best fit models to the \LM\
  (solid) and \LR\ (dashed) data. The models are double power laws
  in $\xi(r)$ mapped to $w(\theta)$ with Limber's equation. The bottom
  panels show the ratio of the \LR\ and \LM\ correlations to the best
  fit \LL\ model. Errorbars are omitted from the top plots for clarity.}
\label{fig:match_cfs}
\end{figure*}

In Fig.~\ref{fig:match_cfs} we show the LRG-LRG and LRG-radio
(hereafter \LL\ and \LR) angular
correlation functions for the three samples we are studying. In the top
panels we show the raw correlations. In
the bottom panels we show the ratio of our measurements
to the best double power-law fit to the \LL\ correlation
function taken from \citet{saw09}.

%\begin{figure*}
%\centerline{\psfig{file=correlations_sdss.ps,width=6cm}\psfig{file=correlations_2slaq.ps,width=6cm,angle=-90}\psfig{file=correlations_aao.ps,width=6cm}}
%\caption{Correlation functions}
%\label{fig:corr_funcs}
%\end{figure*}

In all three samples we study it is apparent that the \LR\
correlation is stronger than the \LL\ autocorrelation, indicating that radio
galaxies are found preferentially in richer environments than their
parent LRG population. The \LR\ cross correlation (like \LL) exhibits
a clear inflection at intermediate scales ($\sim$a few arcmin) which
is generally interpreted as the crossover between the intra and
inter-halo scales. Similar to previous authors
we find that the excess clustering signal from the \LR\ correlation is
strongest in the 1 halo regime. \citet{don09} interpret this as
indicating radio galaxies hold a particularly central
position within their dark matter halo.
At larger scales the difference between the \LL\ and \LR\ correlation
functions is reduced, and becomes negligible above $\sim20$\,arcmin in
the \AAO\ sample. As we discuss later this may be due to the radio
sources being brighter in this higher-redshift sample.

To relate the measured angular correlation function, $w(\theta)$, to
the 3D spatial correlation function, $\xi(r)$, we follow the standard
approach outlined in \citet{phi78}. Given the clear inflection in the
correlation functions shown in Fig.~\ref{fig:match_cfs} we assume the
spatial correlation function can be parametrised as a double power
law and integrate equation~13 from \citet{phi78} to give
$w(\theta)$. In our analysis we fix the break scale in the correlation
function to 1\,$h^{-1}$\,Mpc. We fit for the slopes of the two
power laws and the amplitude of the large-scale
power law. Table~\ref{tab:pl_vals} gives the parameters of fits to the
observed data. 

The large scale slope of $\xi(r)$ in the \LR\ data is essentially
identical in each sample, as it is when compared to the \LL\
correlations. The small 
scale power law is considerably steeper in the radio correlations
indicating that the local environment of radio LRGs is more overdense
compared to the environment external to its own dark matter halo.

\begin{table}
\begin{center}
\caption{Parameters of power-law fits to the \LL\ and \LR\ angular correlation
  functions. The first columns are taken from \citep{saw09}. The fits are
  shown in Fig.~\ref{fig:match_cfs}.}
\begin{tabular}{ccccc}
\hline \hline
 & \multicolumn{2}{c}{\LL} & \multicolumn{2}{c}{\LR} \\
Sample & $r_0$($h^{-1}$\,Mpc) & $\gamma$ & $r_0$($h^{-1}$\,Mpc) & $\gamma$ \\
\hline
SDSS   & 7.35$\pm$0.08 & 2.19$\pm$0.03 & 7.46 & 2.31$\pm$0.07 \\
       & 9.15$\pm$0.16 & 1.85$\pm$0.04 & 9.48$\pm$0.21 & 1.79$\pm$0.08 \\
\hline
2SLAQ  & 6.32$\pm$0.03 & 2.16$\pm$0.01 & 6.27 & 2.32$\pm$0.04 \\
       & 7.78$\pm$0.05 & 1.85$\pm$0.02 & 9.13$\pm$0.17 & 1.82$\pm$0.06 \\
\hline
\AAO\  & 5.96$\pm$0.03 & 2.14$\pm$0.01 & 5.06 & 2.49$\pm$0.05 \\
       & 7.84$\pm$0.04 & 1.81$\pm$0.02 & 8.71$\pm$0.18 & 1.83$\pm$0.07 \\
\hline \hline
\label{tab:pl_vals}
\end{tabular}
\end{center}
\end{table}

\subsection{Radio loud vs. quiet LRG clustering}

Radio galaxies are known to be found preferentially in the most
luminous host galaxies (e.g. \citealt{m+s07}). The most luminous galaxies
tend to also be found in the highest density regions and following
previous authors (e.g. \citealt{wak08a}) we test whether this covariance
is responsible for
the increased \LR\ correlation strength. We construct random
samples of LRGs from our parent catalogues that have the same ($r$-band)
magnitude and $r-i$ colour distributions as the radio LRG catalogues. In
Fig.~\ref{fig:match_cfs} we show the cross correlation between this
matched catalogue and the full LRG sample. We find that they follow
the \LL\ lines almost exactly indicating that the host mass/luminosity
is not the single factor that defines the probability that a LRG is a
radio source, there must be an environmental effect as well.

Our results are in agreement with most previous authors
\citep{wak08a,man09,don09}. \citet{hic09}, on the other hand, finds
that the radio galaxies in his sample cluster with the same amplitude
as galaxies matched in terms of their optical properties. They note
that the result is marginal but also point out that their
radio galaxies have considerably higher stellar
mass to radio power ratios than the other samples, and suggest this as
a potential explanation for the discrepancy. Certainly if the
clustering strength depends on the radio luminosity that could
potentially explain their result.

\subsection{Clustering strength vs. radio luminosity}

%It has been known for some time that radio galaxies are strongly
%clustered with respect to optically selected galaxies, it is less
%clear how 
%the clustering of radio galaxies depends on their radio
%luminosity.
%
\citet{p+p88} found that less luminous FR1 objects tended
to be in overdense regions while more luminous FR2s had a clustering
environment similar to the overall population of radio-quiet
galaxies. However, many authors (e.g. \citealt{p+n91,mag04}) have found no
relationship between radio power and large scale clustering strength. 
\citet{don09} found variations
in the correlation amplitude with radio power that were more pronounced at
smaller spatial scales. At these smaller scales they found that the
correlation amplitude increased with radio power up to
$\sim10^{25}$\,W/Hz, above which the correlation amplitude flattened
and then began to fall again above $\sim10^{25.5}$\,W/Hz.

\begin{figure*}
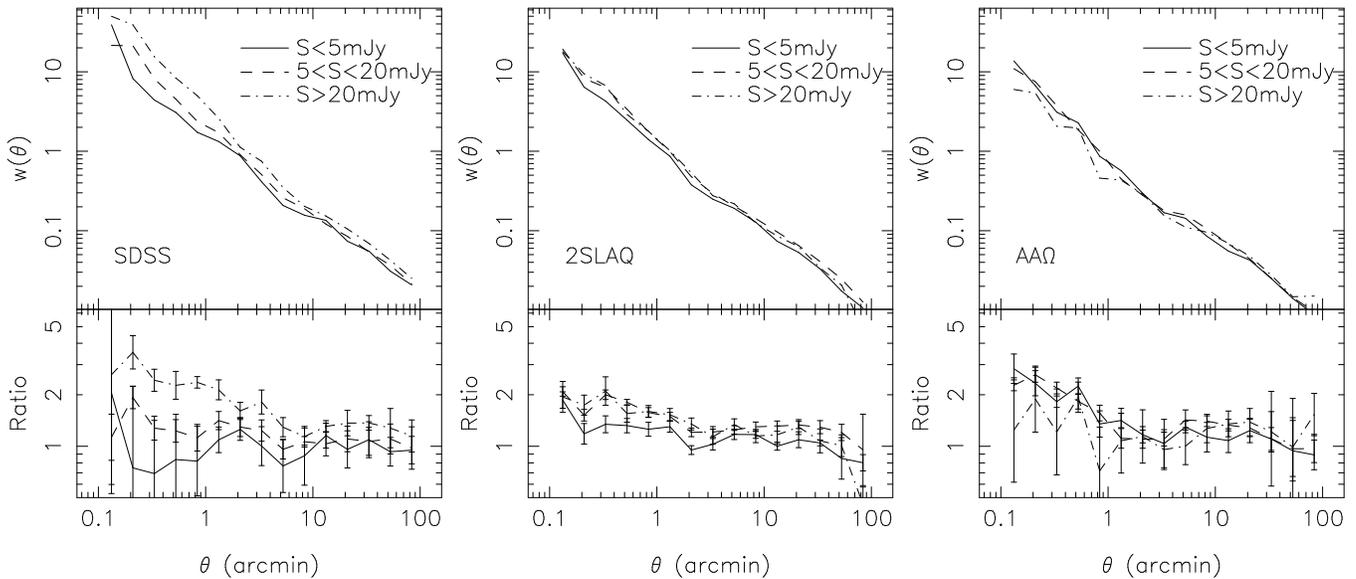

\centerline{\psfig{file=plot_radbin_SDSS.ps,width=5.8cm}\hspace{0.2cm}\psfig{file=plot_radbin_2SLAQ.ps,width=5.8cm}\hspace{0.2cm}\psfig{file=plot_radbin_AAO.ps,width=5.8cm}}
\caption{\LR\ cross correlation functions for the SDSS,
  2SLAQ and \AAO\ samples (left to right) with the radio sample split
  into three bins by radio flux density. The bottom
  panels show the ratio of the correlation functions to the best
  fit \LL\ model. Errorbars are omitted from the top plots for
  clarity. In the SDSS sample there is 
  evidence for a positive correlation between radio flux density and
  the correlation strength. This is not so clear in the other two
  samples.}
\label{fig:radbin_cfs}
\end{figure*}

We divide our sample of radio LRGs into three bins of roughly equal size
by their flux density with cuts at $S_{1.4\,{\rm GHz}}=$5 and 
20\,mJy/Hz. Fig.~\ref{fig:radbin_cfs} shows the \LR\ cross correlations
for each of the flux density bins for the three LRG samples.

We do not find the same behaviour in each LRG sample. In
the SDSS sample there is a clear positive correlation between
radio luminosity and the angular correlation amplitude. As with the
difference between the \LL\ and \LR\ correlations, the luminosity
dependence is
most pronounced at small angular scales ($<$2\,arcmin). In the 2SLAQ
sample the lowest flux density bin is significantly ($>3\sigma$) lower than
the other two subsets that essentially lie on top of each
other. However, in the highest redshift \AAO\ sample we find no
significant difference between the three flux bins.

Our results are broadly consistent with those of \citet{don09}. At the
median redshift of our samples the flux density limits we have chosen
5/20\,mJy correspond to $\log(L_{1.4\,{\rm GHz}}{\rm /(W/Hz)})=$
24.3/24.9, 24.7/25.3 and 25.0/25.5 for the SDSS, 2SLAQ and \AAO\
samples respectively. Hence our results are consistent with the
small-scale radio 
correlation function increasing for radio luminosities
$\lesssim10^{25}$\,W/Hz and then flattening off.

\section{Clustering strength of radio sources at $z<0.68$} \label{sec:evol}

Given our three samples span $z\sim0.68$ to 0.35 ($\sim$2.3\,Gyr of
 cosmic time) we are interested in any signs of evolution in the clustering
strength of radio LRGs. To make a fair
comparison we make our samples as equivalent as possible in terms of
their optical and
radio selection. \citet{saw09} showed that by applying
magnitude limits of $i_{dev}=19.32$ and $20.25$ to the 2SLAQ and \AAO\
samples respectively, they could be made 
roughly equivalent to the SDSS sample in terms of their (optical)
luminosity distribution
and space density (given that the LRG luminosity function does not
evolve strongly over this interval; e.g. \citealt{wak08b}). They
call these optical LRG samples with additional magnitude cuts
SDSS$^*$, 2SLAQ$^*$ and \AAO$^*$ . In
addition, we increase the radio flux limit of the
SDSS$^*$ and 2SLAQ$^*$ samples to 13.8 and 4.9\,mJy respectively, in
order to be equivalent to a radio 
power of $\sim10^{24.72}$\,W/Hz at the median redshift of each
sample. This results in 3576, 4755, 5089 objects in our radio matched
samples we will call SDSS\rs, 2SLAQ\rs\ and \AAO\rs.

\begin{figure}
\centerline{\psfig{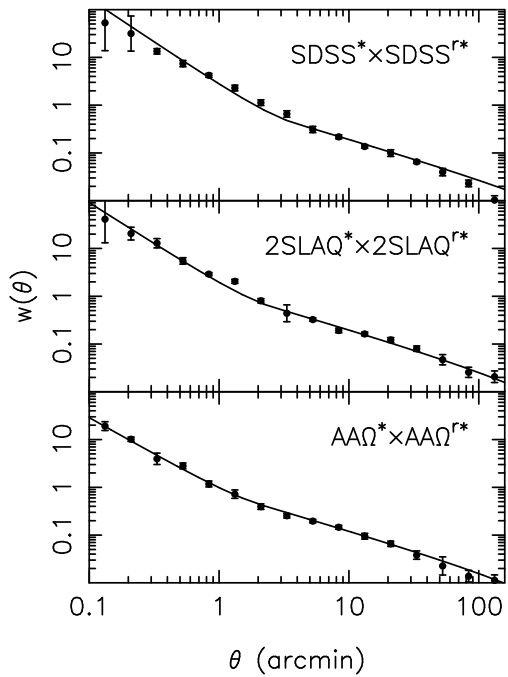}}
\caption{The SDSS$^*\times$SDSS\rs,
  2SLAQ$^*\times$2SLAQ\rs\ and \AAO$^*\times$\AAO\rs\ cross
  correlation functions. The solid lines show the best-fit double
  power laws in  $\xi(r)$ mapped to $w(\theta)$ with Limber's equation.}
\label{fig:3s_cfs_lr}
\end{figure}

In Fig.~\ref{fig:3s_cfs_lr} we show the cross correlations between
the radio matched and optically matched samples along with a best-fit
double power law in $\xi(r)$ mapped to $w(\theta)$ with Limber's
equation. Again 
the break in the power law is assumed to be at $1h^{-1}$\,Mpc 
when performing the fit, in addition to which we fix the slope of the
large-scale power law to be $\gamma=1.8$ to facilitate comparisons of the
clustering amplitude.

\begin{figure}
\centerline{\psfig{file=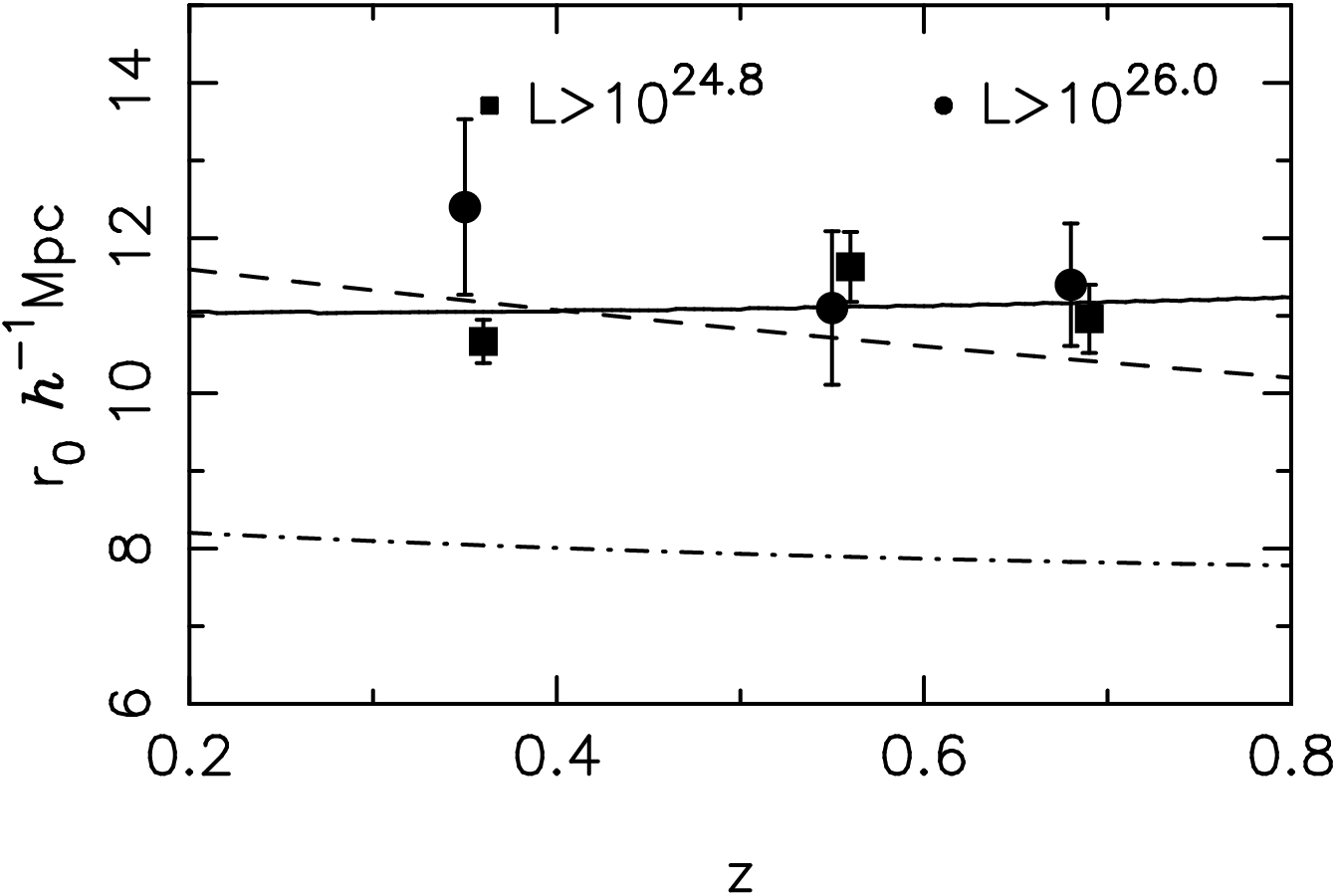,width=7cm,angle=-90}}
\caption{The large-scale cross-correlation amplitude for the
  SDSS$^*\times$SDSS\rs, 
  2SLAQ$^*\times$2SLAQ\rs\ and \AAO$^*\times$\AAO\rs\ correlations
  (squares) and the SDSS$^*\times$SDSS\ts,
  2SLAQ$^*\times$2SLAQ\ts\ and \AAO$^*\times$\AAOts correlations
  (circles; these have been offset horizontally for clarity). Neither
  show any evidence for a trend with redshift. The lines show
  models for the cross-correlation amplitude based on differing
  assumptions about the radio-source autocorrelation. The solid and
  dashed lines are constant dark halo mass and long lived models
  respectively, fit to the lower luminosity (square) points. The
  dot-dashed line assumes that the radio autocorrelation is the same
  as that measured for quasars.}
% The lines show two
%  models for the cross-correlation amplitude based on differing
%  assumptions about the radio-source autocorrelation. The dashed line
%  assumes a long-lived clustering model and is fit to the $L>10^{24}$\,W/Hz
%  points. The solid line assumes a constant dark halo mass of
%  $3\times10^{12}$\,M$_\odot$ (see text) for the autocorrelation and
%  has no free parameters. In both cases the cross-correlation amplitude
%  is derrived assuming the best-fit long-lived model from
%  \citet{saw09} and using $\xi_{11}\xi{22}=\xi{12}^2$.}
\label{fig:r0_lr_evol}
\end{figure}

%The primary goal of this paper is to investigate the evolution in the
%clustering strength of radio galaxies and in 
Fig.~\ref{fig:r0_lr_evol} shows
the large-scale
($>1h^{-1}$\,Mpc) clustering amplitude as a function of
redshift (squares). The results are consistent with a constant clustering
amplitude with little evidence for evolution.

\begin{figure}
\centerline{\psfig{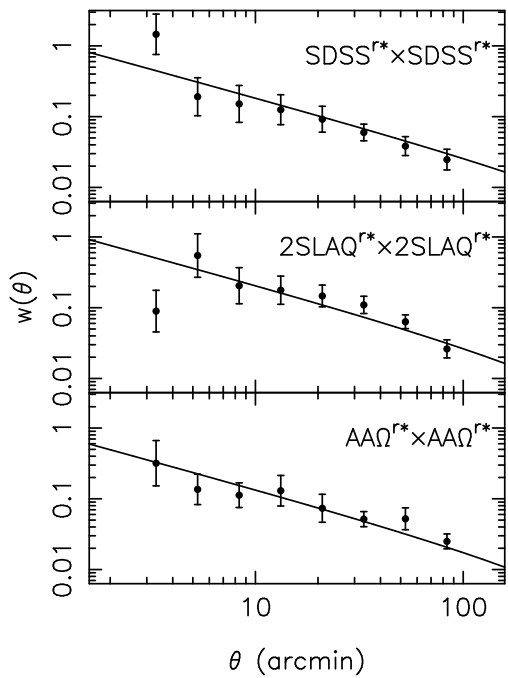}}
\caption{Autocorrelation functions for the three samples after they
  have been made approximately equivalent in terms of their optical and
  radio selection. The solid line gives the best fit relation in terms
  of $\xi(r)=(r_0/r)^\gamma$ propagated through Limber's formula.}
\label{fig:3s_cfs}
\end{figure}

To make a comparison with previous work on the clustering of radio
sources we calculate the radio LRG two-point
autocorrelation function in each sample and show the results in
Fig.~\ref{fig:3s_cfs}. The lines in the figure show the best fit
power law in $\xi(r)$ ($\gamma$ fixed to 1.8) mapped to $w(\theta)$
with Limber's equation. Table~\ref{tab:pl_vals_match} summarises the results
of our power law fitting in this section.

%To compare the clustering strengths of the three
%samples we calculate
%\begin{equation}
%\xi_{20}=\frac{3}{20^3}\int_0^{20}\xi(r)\,r^2\,{\rm d}r
%\end{equation}
%and plot the values for each sample in Fig.~\ref{fig:3s_int}.
%
%
%\begin{figure}
%\centerline{\psfig{file=plot_int_evol.ps,width=7cm}}
%\caption{}
%\label{fig:3s_int}
%\end{figure}
%
%It is clear that if there is any evolution it is towards a weaker
%clustering strength at lower redshifts, although this is weak. This
%may be of no surprise since the
%(non-radio matched) parent samples showed a similar lack of
%evolution in clustering strength. However, while the LRG luminosity
%function and clustering strength does not evolve strongly for $z<0.68$
%\cite{saw09,wak08b}, the bright end of the radio luminosity function
%increases by a factor of $\sim10$ from $z=3.5$ to 6.8
%\citep{d+p90,sad07}. Our results indicate that, while more LRGs are
%radio sources at $z=0.68$ compared to $z=3.5$ the environments
%of these obects do not change significantly over the same interval.

\citet{bra05} measured the clustering strength of low-luminosity radio
galaxies at $z\sim0.3$ and found $r_0=6.1\pm1.1\,h^{-1}$\,Mpc assuming
$\xi(r)=(r/r_0)^{-1.8}$. This is significantly
($\sim3.1\,\sigma$) less than the value of
$r_0=11.0\pm1.2\,h^{-1}$\,Mpc found by \citet{p+n91} at
$z<0.1$. \citet{bra05} explained some of the discrepancy by assuming
the clustering amplitude would increase accordingly with linear theory
from $z=0.3$ to 0. \citet{wak08a} derived an autocorrelation
clustering strength (extrapolated from the
cross correlation) of $r_0=12.3\pm1.2\,h^{-1}$\,Mpc for radio sources
in the 2SLAQ sample.
%Fixing the slope of $\xi(r)$ to -1.8 we find
%clustering amplitudes of
%12.4$\pm1.6$, 11.1$\pm1.3$ and 11.4$\pm1.3\,h^{-1}$\,Mpc
%%8.4$\pm1.6$, 11.3$\pm1.3$ and 8.2$\pm1.3\,h^{-1}$\,Mpc
%for the SDSS\rs, 2SLAQ\rs\ and \AAO\rs\ samples respectively.

\begin{figure}
\centerline{\psfig{file=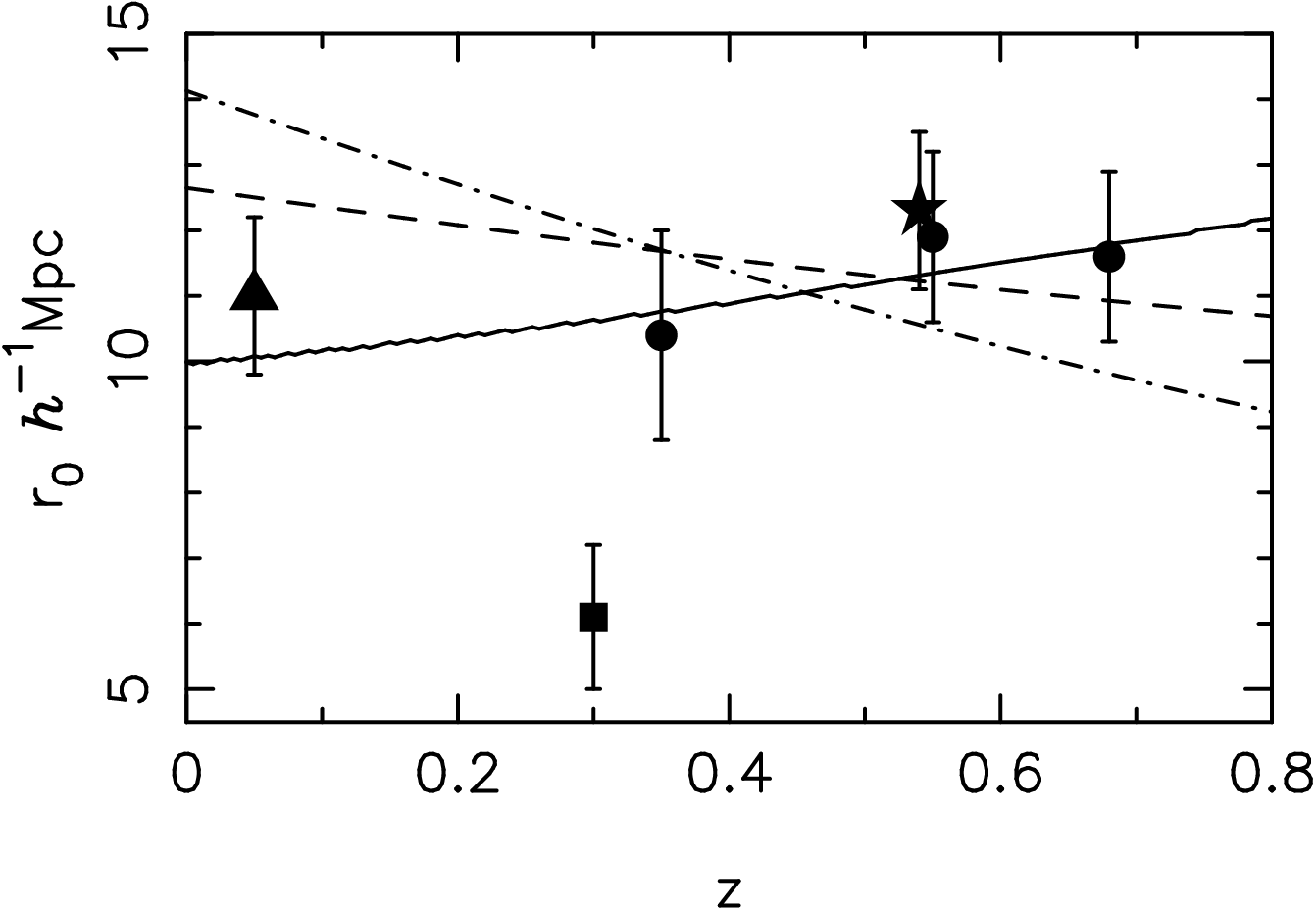,width=7cm,angle=-90}}
\caption{The radio galaxy$\times$radio galaxy autocorrelation amplitude
  $r_0$ as a function of
  redshift. The circles correspond to our three samples in
  addition to points from \citet{p+n91} (triangle), \citet{bra05}
  (square) and \citet{wak08a} (star; this point is offset horizontally
  for clarity). The measurements are consistent with
  no evolution in the clustering amplitude with redshift up to
  $z\sim0.68$. The \citet{bra05} point is significantly offset from
  the rest, potentially due to their sample being dominated by
  fainter objects. The lines give three model
  fits to the data excluding the \citet{bra05} point. The models are
  constant dark halo mass (solid), long-lived (dashed) and linear
  growth (dot-dashed).}
\label{fig:r0evol}
\end{figure}

In Fig~\ref{fig:r0evol} we show the values for $r_0$ calculated in this
paper (circles) along with the values found by \citet{p+n91}
(triangle), \citet{bra05} (square) and \citet{wak08a} (star). A best
fit to these points gives a gradient consistent with zero
($<0.5\sigma$).

The flux limits of the \citet{p+n91} and \citep{wak08a} samples are
such that they sample $<2$ times fainter in terms of radio
luminosity at their median redshifts compared to our samples. Hence, these
surveys are sampling relatively bright objects 
where we do not find clustering strength correlating with radio power
(Fig.~\ref{fig:radbin_cfs}). The \citet{bra05} sample on the other
hand goes $\sim6.5$ times fainter. This difference may 
explain why their result is lower than the other values.

%into the regime where clustering
%strength is correlated with the radio luminosity of the galaxies. This
%may explain why their result is somewhat lower than the other values.

%Of cource between $z=0.68$ and 3.8 the clustering of dark matter
%increases significantly indicateing that the bias of these radio
%galaxies does change with redshift.

\begin{table*}
\begin{center}
\caption{Parameters of power law fits to the angular correlation
  function for samples matched in terms of their optical and radio
  properties. Samples included are denoted $*$ (optically selected LRGs),
  $r*$ (radio LRGs with $L_{1.4\text{GHz}}>10^{24.72}$\,W/Hz) and $26*$
  (radio LRGs with $L_{1.4\text{GHz}}>10^{26}$\,W/Hz).}
\begin{tabular}{ccccccc}
\hline \hline
 & \multicolumn{2}{c}{SDSS} & \multicolumn{2}{c}{2SLAQ} &
 \multicolumn{2}{c}{\AAO} \\
Correlation & $r_0$($h^{-1}$\,Mpc) & $\gamma$ & $r_0$($h^{-1}$\,Mpc) &
 $\gamma$ & $r_0$($h^{-1}$\,Mpc) & $\gamma$ \\
\hline
$*\times r*$  & 4.55  & 2.81$\pm$0.09 & 5.07  & 2.72$\pm$0.08 & 5.49
 & 2.53$\pm$0.06 \\ 
              & 10.67$\pm$0.28 & 1.8  & 11.63$\pm$0.45 & 1.8  &
 10.96$\pm$0.44 & 1.8  \\ 
\hline
$r*\times r*$  & 10.4$\pm$1.6 & 1.8 & 11.9$\pm$1.3 & 1.8 &
 11.6$\pm$1.3 & 1.8 \\ 
\hline
$*\times 26*$ & 10.67$\pm$1.13 & 1.8 & 11.63$\pm$0.99 & 1.8 &
 10.96$\pm$0.79 & 1.8 \\ 
\hline \hline
\label{tab:pl_vals_match}
\end{tabular}
\end{center}
\end{table*}

\subsection{Evolution of the clustering amplitude of radio galaxies}

%In Fig.~\ref{fig:r0evol} we fit a series of models for the evolution
%of the clustering amplitude of radio sources.

It is clear from Fig.~\ref{fig:r0evol} and the squares in
Fig.~\ref{fig:r0_lr_evol} that there is little evidence for evolution
in the clustering amplitude of either the auto or cross
correlation with redshift. Fitting $r_0\propto(1+z)^\alpha$ to the
measurements in Fig.~\ref{fig:r0evol} (excluding the
\citealt{bra05} point) gives $\alpha=0.20\pm0.30$. The same fit for the
cross correlation (squares in Fig.~\ref{fig:r0_lr_evol}) gives
$\alpha=0.23\pm0.21$. In addition we fit a series of physically
motivated models to our data:

\begin{enumerate}
\item{In linear theory structures evolve as $\xi_m\propto D^2(z)$ where 
$D(z)$ is the growth factor and $\xi_m$ is the matter correlation
function. Structures that grow according to this 
have a constant bias ($b$) where $b^2=\xi/\xi_m$.
In Fig.~\ref{fig:r0evol} we fit a constant bias model to our data
and plot the results as the dot-dashed line. This
model is only marginally consistent with our data ($\chi^2=11.11;
P(\chi^2,\nu=4)\sim2.5$\,\%).}

\item{
%Models that involve less evolution are better
%fits to our data.
We further include the long-lived model of \citet{fry96}
that assumes no change in the comoving number density of galaxies and
that clustering grows solely due to gravitational effects. The long
lived model has been shown to describe evolution in the LRG$\times$LRG
correlation function \citep{saw09}. Fits to the radio cross and
auto correlation amplitudes are shown as dashed lines in
Figs.~\ref{fig:r0_lr_evol} and~\ref{fig:r0evol}.
Note that the long-lived model does not require that radio sources
themselves exist for cosmological timescales. Rather that the dark
matter halos that host them evolve passively under the influence of
gravity. This model involves little evolution and hence gives a better fit to
our data.} 

\item{We also fit a model in which dark halo mass is held constant with
redshift (solid lines Figs.~\ref{fig:r0_lr_evol} and~\ref{fig:r0evol};
see \citealt{saw09} for our formalism for relating bias to 
dark halo mass). This type of model has been shown to fit the evolution
of the quasar autocorrelation function over $0\lesssim z \lesssim2.5$.}
\end{enumerate}
%We also fit a a constant dark
%halo mass model that has been shown to describe the evolution of the quasar
%autocorrelation function \citep{croom05,ros09} (solid lines
%Figs.~\ref{fig:r0_lr_evol} and~\ref{fig:r0evol}).

To relate our models for the radio auto correlation to the cross
correlation amplitudes in Fig.~\ref{fig:r0_lr_evol} we
assume the LRG clustering follows the best-fit long-lived model
from \citet{saw09}. We then assume that the cross-correlation of two samples
($\xi_{12}$) is related to the two autocorrelations ($\xi_{11}$ and
$\xi_{22}$) by (e.g. \citealt{wak08a})
\begin{equation}
\xi_{12}^2=\xi_{11}\xi_{22}.
\label{equ:a_x}
\end{equation}

The long-lived and constant mass models fit our data acceptably
($P(\chi^2,\nu)>10$\,\%) with the exception of the long-lived model
fit to the cross correlation results. In this case we find
$\chi^2=9.08$ with $P(\chi^2,\nu=2)\sim1.1$\,\% indicating that the
long-lived model may not accurately describe radio source clustering.
%With the exception of the long-lived model fit to the cross-correlation
%points, the models are consistent
%($P(\chi^2,\nu)>10$\,\%) with both
%the auto correlations (all points in Fig.~\ref{fig:r0evol} excluding
%\citep{bra05}, and the cross correlations (squares in
%Fig.~\ref{fig:r0_lr_evol}). In the case of the long-lived model fit to
%the cross-correlation data we find $\chi^2=9.08$ with
%$P(\chi^2,\nu=2)\sim1.1$\,\% indicating that the long-lived model may
%not accurately describe radio source clustering.
The dark halo masses derived for the
radio sources are $9.4$ and
$5.6\times10^{13}h^{-1}$\,M$_\odot$ for the cross and auto
correlation respectively.

The offset in the derived dark halo masses may be due to
equation~\ref{equ:a_x} not holding in our data. In this paper and
\citet{saw09} large-scale correlation amplitudes have been calculated
for the $*\times*$ LRG auto correlation, $*\times*r$ cross correlation
and the $*r\times*r$ autocorrelation. Calculating the cross
correlation amplitude from the autocorrelations and
equation~\ref{equ:a_x} gives $r_0=9.75\pm0.75$, $10.28\pm0.56$ and
$10.26\pm0.58h^{-1}$\,Mpc for the SDSS, 2SLAQ and \AAO\ samples
respectively. These values are consistently lower than the amplitudes
measured from the cross correlation, although only by $\sim1$ to
$2\sigma$. Taking the three samples together the offset is
$\sim3\sigma$. The offset may be due to two causes. First, the
fits are not completely equivalent; we fit double power laws over a
larger range
of $\theta$ for the $*\times*$ and $*\times*r$ correlations. Second,
equation~\ref{equ:a_x} assumes a linear bias model that may not hold
in this case.

\section{Clustering of luminous radio sources at $z<0.68$} \label{sec:Levol}

The SDSS\rs, 2SLAQ\rs\ and \AAO\rs\ samples
are dominated by objects below the $\sim10^{26}$\,W/Hz divide between
brighter FRI and FRII sources. $10^{26}$\,W/Hz is also roughly the
divide between radio AGN host galaxies that show no emission lines and
those that do. Furthermore, above $\sim10^{26}$\,W/Hz radio sources show
strong number density evolution, more 
similar to quasars. The clustering strength of quasars evolves weakly
with redshift \citep{croom05,ros09}. In this section we
use our sample to test for evolution in the clustering amplitude of
bright radio LRGs.

\begin{figure}
\centerline{\psfig{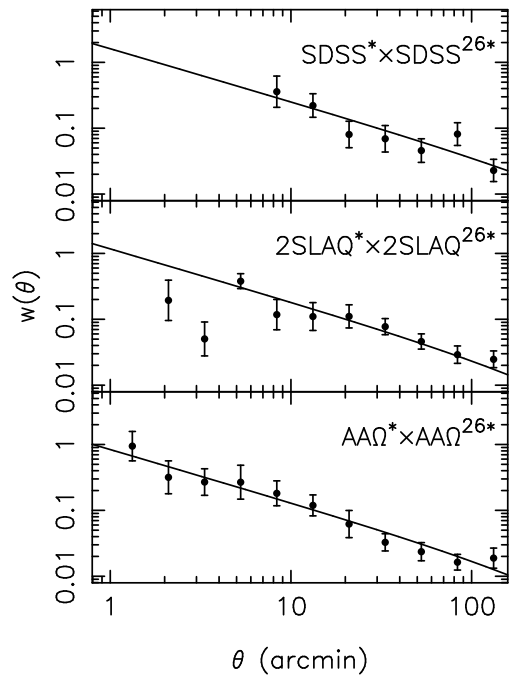}}
\caption{The SDSS$^*\times$SDSS\ts,
  2SLAQ$^*\times$2SLAQ\ts\ and \AAO$^*\times$\AAOts\ cross
  correlation functions. The solid line shows the best fit solutions
  to $\xi(r)=(r_0/r)^{1.8}$ propagated through Limber's formula.}
\label{fig:star26}
\end{figure}

A radio luminosity of $10^{26}$\,W/Hz corresponds to flux limits  of
$\sim266$, 95 and 57\,mJy at $z\sim0.35$, 0.55 and 0.68
respectively. Cutting back our samples to these limits leave 49, 224
and 283 objects in what we will call the SDSS\ts, 2SLAQ\ts\ and
\AAOts\ samples. These are not large enough to calculate an
autocorrelation amplitude so we calculate the two-point cross
correlation with the corresponding (starred) LRG
catalogues. Fig.~\ref{fig:star26} shows the cross correlations along
with best fit solutions to $\xi(r)=(r_0/r)^{1.8}$ propagated through
Limber's formula. In Fig.~\ref{fig:r0_lr_evol} the circles show $r_0$
for these bright objects as a function of redshift.
%Again we find no evidence for evolution in the clustering
%amplitude.

Note that while lower-luminosity radio AGN are
typically found in LRGs with red stellar spectra, the brightest
sources can be found in both broad and narrow emission-line
objects that have younger stellar populations (e.g. \citealt{joh08}).
Hence, while an LRG selection finds most
hosts of radio galaxies in the range $10^{23}<L<10^{26}$\,W/Hz, above this
there is a significant population of blue objects that
may be missed by the optical selection.

%Again we find no evidence for evolution in the clustering
%amplitude. This does not necessarily mean that the clustering
%amplitude of the brightest radio sources does not evolve in a similar
%fashion to quasars since the clustering strength of quasars evolves weakly
%in the redshift interval 0.68 to 0.35.

\subsection{Evolution of the clustering amplitude of luminous radio sources}

As with the less luminous sample, we find very little evidence for
evolution in the clustering amplitude of these brightest radio sources.
Fitting the cross correlation points (circles
Fig.~\ref{fig:r0_lr_evol}) for
$r_0\propto(1+z)^\alpha$ gives $\alpha=-0.38\pm0.52$.

We also compare our results for the clustering of the brightest radio
sources with that of quasars. To do this we
assume all quasars cluster such that their derived dark halo mass is
$3\times10^{12}$\,M$_\odot$ at all redshifts \citep{croom05,ros09} and
work backwards to derive the clustering strength as a
function of redshift. We then convert this to a cross-correlation
clustering strength making the assumptions discussed above.
%assuming $\xi_{11}\xi{22}=\xi{12}^2$
%(e.g. \citealt{wak08a}) and the best-fit long-lived model \citep{fry96}
%for LRG$\times$LRG clustering strength from \citet{saw09}. 
We plot the
derived value for the cross-correlation amplitude in
Fig.~\ref{fig:r0_lr_evol} as the dot-dashed line.

While our derivation is approximate it suggests that the clustering
amplitude of 
bright radio sources may evolve slowly with redshift and in an almost
luminosity-independent manner, similar to the results found for quasars.
There is also strong evidence that the objects in our sample cluster
considerably more strongly than quasars, as previously noted for fainter
radio sources. The implication is that these bright radio sources must
inhabit considerably denser regions of space and correspondingly more
massive dark matter halos compared to quasars. There is less difference
between the large-scale clustering environment of luminous and faint radio
sources. However, the small scale $r<1$h$^{-1}$Mpc clustering environment
of the brightest radio sources appears significantly denser than the
environment of the faintest sources. This may indicate the effect of
the radio source on its immediate surroundings, given that the
larger-scale environment is luminosity independent.

\section{Conclusions}

We have cross-matched three large photometric samples of LRGs to the FIRST
and NVSS surveys to define three samples of radio LRGs. We have then
measured the 2-point correlation function to investigate the clustering
properties of the samples. Similar to previous authors we find
evidence that: 1) radio LRGs are more strongly clustered than non-radio
galaxies matched in terms of luminosity and colour; 2) the
clustering strength of radio LRGs at $r<1$h$^{-1}$Mpc increases with radio
luminosity up to $L\sim 10^{25}$ W/Hz and then remains roughly constant but
at $r>1$h$^{-1}$Mpc the clustering is independent of radio luminosity. We
further find that inconsistencies between the radio-LRG cross-correlation
and auto-correlation amplitudes may suggest that a simple linear bias
model may be insufficient to describe the relation between radio-LRG
and/or LRG clustering and mass clustering.

However, the primary goal of this work was to investigate any
evolution in the clustering strength of radio galaxies with redshift. We
find no evidence for evolution in the large scale ($r>1$h$^{-1}$Mpc)
clustering amplitude of radio-LRGs. We show that our radio $\times$
radio LRG autocorrelations are consistent with previous authors
indicating no evolution in comoving coordinates in the redshift range
$0<z<0.68$ ($\sim6$Gyr). Furthermore, we make use of cross-correlations
to increase the precision of our results and still find no evidence for
evolution. We then restrict our samples to objects with $L >
10^{26}$W/Hz and again find no evidence for evolution in the correlation
amplitude. Our results are consistent with a single dark halo mass of
$9\times10^{13}h^{-1}$\,M$_\odot$ for all radio LRGs in the redshift
range $0<z<0.68$. This could be compared to  quasars that appear to
inhabit halos of $m_{DH}=3\times10^{12}h^{-1}$\,M$_\odot$ again
reasonably independent
of redshift and luminosity. The significantly more massive halos for the
high-luminosity radio galaxies, at least,  may provide a problem for
unified AGN models. Finally, our radio-LRG cross-correlations are
inconsistent with a model in which LRG clustering follows a long-lived
model and radio sources are randomly sampling the LRGs.

\bibliography{bib}

\end{document}